# Pharmacoprint – a combination of pharmacophore fingerprint and artificial intelligence as a tool for computer-aided drug design

*Dawid Warszycki*[a], *Łukasz Struski*[b], *Marek Śmieja*[b], *Rafał Kafel*[a], *Rafał Kurczab*[a,*]

[a]Maj Institute of Pharmacology Polish Academy of Sciences, Smetna 12 Street, 31-343 Cracow, Poland

[b]Faculty of Mathematics and Computer Science, Jagiellonian University, 6 Lojasiewicza Street, 30-348, Cracow, Poland

**Abstract**

Structural fingerprints and pharmacophore modelling are methodologies that have been used for at least two decades in various fields of cheminformatics: from similarity searching to machine learning (ML). Advances in *in silico* techniques consequently led to combining both these methodologies into a new approach known as pharmacophore fingerprint. Herein, we propose a high-resolution, pharmacophore fingerprint called Pharmacoprint that encodes the presence, types, and relationships between pharmacophore features of a molecule. Pharmacoprint was evaluated in classification experiments by using ML algorithms (logistic regression, support vector machines, linear support vector machines, and neural networks) and outperformed other popular molecular fingerprints (i.e., Estate, MACCS, PubChem, Substructure, Klekotha-Roth, CDK, Extended, and GraphOnly) and ChemAxon Pharmacophoric Features fingerprint. Pharmacoprint consisted of 39973 bits; several methods were applied for dimensionality reduction, and the best algorithm not only reduced the length of bit string but also improved the efficiency of ML tests. Further optimization allowed to define the best parameter settings for using Pharmacoprint in discrimination tests and for maximizing statistical parameters. Finally, Pharmacoprint generated for 3D structures with defined hydrogens as input data was applied to neural networks with a supervised autoencoder for selecting the most important bits and allowed to maximize Matthews Correlation Coefficient up to 0.962. The results show the potential of Pharmacoprint as a new, perspective tool for computer-aided drug design.

## Introduction

Currently, fingerprints have become one of the most popular forms of encoding the structure of a chemical compound in an understandable way for computers and have found wide applications in cheminformatics. The idea underlying molecular fingerprint is to apply a function to the molecule to generate a bit vector or less frequent count vector. Easy application of fingerprints in tasks such as similarity searching, clustering, and classification problems have made them an essential tool in computer-aided drug design[1,2,11–18,3–10]. Fingerprints can be divided into two groups: in the first group, each bit describes a precisely defined structural pattern (non-hashed fingerprint, e.g., Klekotha-Roth fingerprint), while in the second group, there is no assigned bit meaning (hashed fingerprint, e.g., GraphOnly fingerprint).

Another commonly used *in silico* method is pharmacophore modeling. A pharmacophore is the spatial orientation of the various structural features of a molecule responsible for interaction with a biological target and for triggering the biological effect.[19] Pharmacophore is a commonly used technique in virtual screening campaigns and other research studies[8,20–26]. The pharmacophore filter itself can be used as a part of screening cascade[25,26] or even as a standalone tool[8,27].

Fingerprints as well often play the role of input for machine learning (ML) methods; thus, creating a fingerprint that extensively describes the pharmacophore properties of a chemical structure will extend the application area of this methodology.

The idea of combining fingerprint and the pharmacophore model is not new, and this concept has already been reported in the literature[28–31]; however, despite the creation of high-resolution representations,[32] the topic leaves much room for further exploration. One of the oldest approaches is proposed by McGregor and Muskal, who used 10549 bits signature to describe the three-point pharmacophores.[31] Wood et al. proposed the inclusion of a four-point pharmacophore in another pharmacophore fingerprint representation, but it extended fingerprint to more than 300000 bits, which made its application extremely time- and CPU-consuming.[29] ChemAxon introduced an atom-pair-based 2D pharmacophore fingerprint in its software. This fingerprint is defined as the collection of all atom-atom pharmacophore feature pairs along with their topological distances.[33] Nevertheless, this signature cannot be translated into a binary vector.

In the present study, the concept and construction of a pharmacophore fingerprint (hereafter called Pharmacoprint) that describes the pharmacophore features in a binary form are presented. The application of the fingerprint as input data for ML experiments is considered and evaluated (Figure 1). Pharmacoprint showed better performance than other commonly used binary fingerprints. Unfortunately, Pharmacoprint cannot be compared with most of previously listed pharmacophore fingerprinting approaches due to (i) their structure-based nature,[29] (ii) availability only in commercial software,[30] (iii) non-availability,[31] (iv) not standardized length,[34] and output in non-binary form.[33] Furthermore, to speed up the processing time of Pharmacoprint and to compress the representation, a new methodology based on deep neural networks was designed to reduce this signature to a 100-dimensional representation, which was sufficient to obtain comparable performance. This new and robust approach (available in a form of free-available ready-to-use python script utilizing RDKit libraries) can boost computer-aided drug design and find broad application in *in silico* studies.

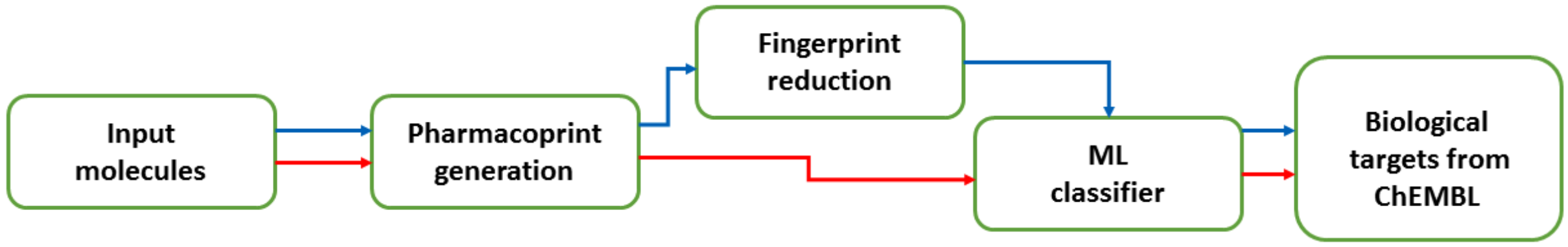

**Figure 1.** General scheme of experiments performed within the study. The blue arrows show the experimental setting where different algorithms were used to reduce the Pharmacoprint dimension, whereas, the red arrows illustrate the experimental path without reduction of the Pharmacoprint. Details are described in the Methodology section.

## Methodology

### *Development of Pharmacoprint*

The main role of the algorithm is to identify pharmacophore features within a molecule and determine inner-feature distances measured as the topological distance in bond as a unit. Feature definitions and distance bins are user-definable. Pharmacoprint uses RDKit libraries and enables fast and easy utilization in a form of a free-available python script.

To illustrate how the algorithm works, let's assume that we have only two pharmacophore features (A and B; Figure 2), all possible combinations of two or three features, and two distance bins (2 or fewer bonds and more than 2). The two pharmacophore features (A and B) can be combined in three pairs (AA, AB, and BB) and four triplets (AAA, AAB, ABB, and BBB). Distance between the pair of features can be considered as within the first bin (2 or fewer bonds; bin '0') or in the second bin (more than 2 bonds; bin '1'), which implies that a single, two-point pharmacophore is described by two bits. Three-point pharmacophores are described by eight bits because three inner-distances, which can be within the first or second distance bin, can generate 8 possible solutions. The entire signature for such a representation consists of 38 bits.

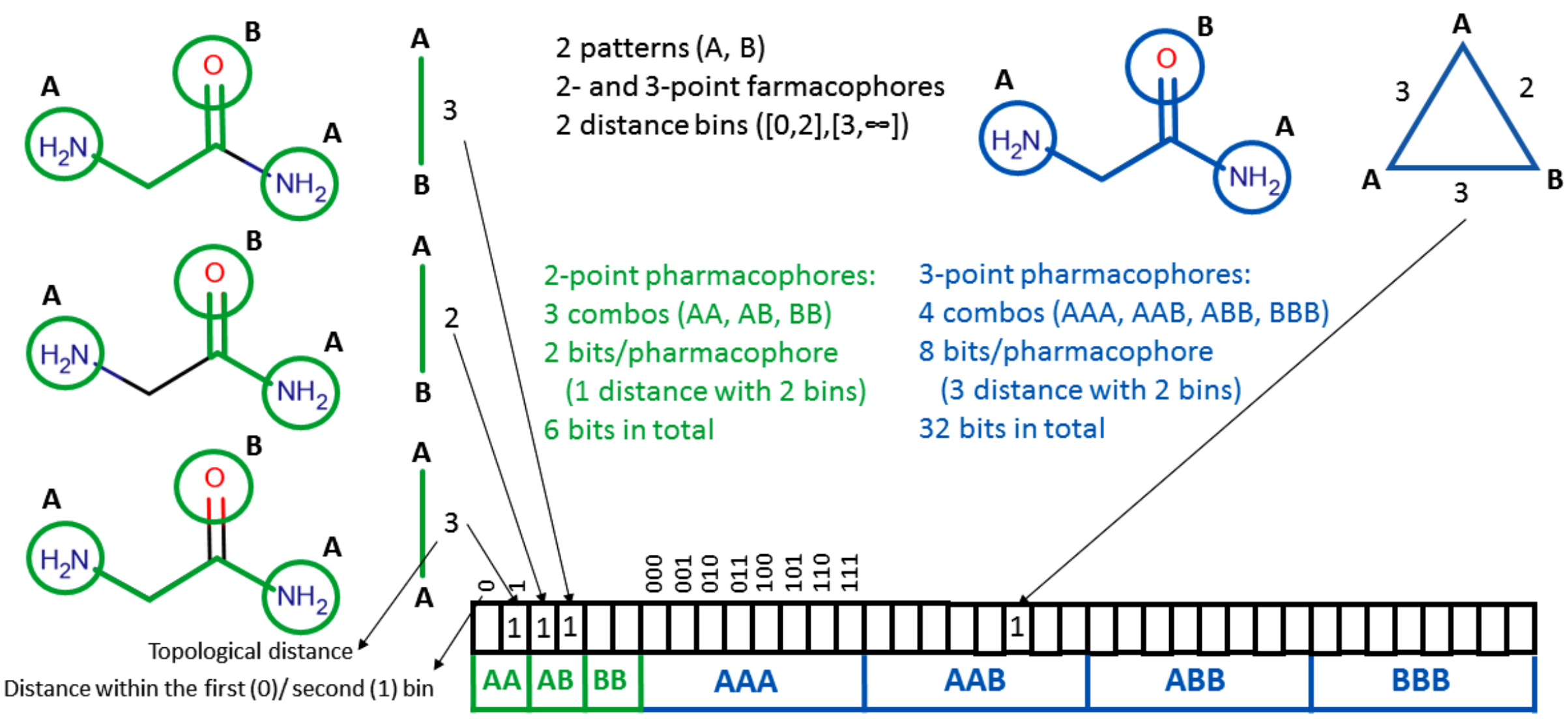

**Figure 2.** Pharmacoprint generation scheme where two features, 2-point and 3-point pharmacophores, and 2 distance bins define 38 bits signature. All possible combinations of pharmacophore features are generated (AA, AB, BB, AAA, AAB, ABB, and BBB). Two bits per pair of pharmacophore features and eight bits per three pharmacophore features describe all possible distance range combinations between particular features.

In the present study, to test the performance of Pharmacoprint, we used the following settings: the eight pharmacophore feature types originally listed by Gobbi and Poppinger (hydrogen bond acceptor, hydrogen bond donor, basic group, acidic group, hydrophobic group, halogen, attachment point to an aliphatic ring, and attachment point to an aromatic ring)[35] and seven distance bins ((<1,3), (<3,4), (<4,5), (<5,6), (<6,7), (<7,8), (<8,100), all distances measured in bond as a unit). A combination of eight pharmacophore features, seven distance bins, and all 2- and 3-point pharmacophores resulted in a 39973-bit long signature.

### *Compound datasets*

To perform a broad range of tests, 15 protein targets were selected from the ChEMBL database (version 26),[36] which ensured the diversity of both the protein targets and the structures of the active compounds (Table 1). As ChEMBL contains numerical values of particular parameters that determine the activity of the compounds, only molecules whose activities were quantified by $K_i$, p$K_i$, or $IC_{50}$ and have been tested in human protein assays (except HIV targets) were considered. The $IC_{50}$ values were recalculated to $K_i$ by using the following expression: $K_i = IC_{50}/2$ (the conversion factor of 2 was suggested by Kalliokoski et al. [37]). Fetched compounds were separated into actives (p$K_i$ or equivalent > 7) and inactives (p$K_i$ or equivalent < 6) based on previously utilized approaches.[22,38,39] For every single set of actives, a collection of putative inactives from the ZINC database were selected in a ratio of 9 inactives per 1 active (Table 1.);

however, this set was only used for comparison of different types of fingerprints (see the first paragraph in the Results section).[40] In the remaining experiments, only a set of true inactives were used.

**Table 1.** List of the receptors used in the present study with the number of appropriate actives, inactives, and compounds from ZINC

| Receptor | Target ChEMBLID | Abbreviation | Actives | Inactives | ZINC |
|---|---|---|---|---|---|
| Serotonin 5-$HT_{2A}$ | CHEMBL224 | 5-HT2A | 2165 | 623 | 19485 |
| Serotonin 5-$HT_{2C}$ | CHEMBL225 | 5-HT2C | 1150 | 574 | 10350 |
| Serotonin 5-$HT_6$ | CHEMBL3371 | 5-HT6 | 2076 | 299 | 18684 |
| Dopamine $D_2$ | CHEMBL217 | D2 | 2422 | 1299 | 21798 |
| HIV integrase | CHEMBL3471 | HIVint | 108 | 690 | 972 |
| HIV protease | CHEMBL243 | HIVprot | 2999 | 809 | 26991 |
| NMDA receptor | CHEMBL330 | NMDA | 125 | 121 | 2970 |
| Nociceptin opioid receptor | CHEMBL2014 | NOP | 183 | 249 | 1647 |
| cathepsin B | CHEMBL4072 | catB | 576 | 608 | 5184 |
| cathepsin L | CHEMBL3837 | catL | 244 | 662 | 2196 |
| Delta opioid receptor | CHEMBL236 | delta | 1772 | 908 | 15948 |
| Kappa opioid receptor | CHEMBL237 | kappa | 1820 | 589 | 16380 |
| Mi opioid receptor | CHEMBL233 | mi | 2208 | 803 | 19872 |
| NPC cholesterol transporter | CHEMBL1293277 | NPC1 | 104 | 16152 | 936 |
| HIV reverse transcriptase | CHEMBL247 | HIVrev | 544 | 698 | 4896 |

***Tested fingerprints***

For all compounds in the datasets, nine types of fingerprints were generated by PaDEL-Descriptor software (Table 2)[41]. Three of the used fingerprints applied hashing function (CDK,

Extended, and GraphOnly), while the remaining fingerprints (except Pharmacoprint) used structural keys as a definition for each bit. Moreover, as it was mentioned in the introduction section, there is no free-available method of encoding pharmacophore features in the form of a bit string so far. However, to compare Pharmacoprint with another pharmacophore fingerprint we transformed ChemAxon PF (pharmacophore features) fingerprint to binary form.[33] This representation was converted to pharmacophore feature pair histograms, which were merged into one signature, where values greater than one were presented as a bit with a value of '1'. Finally, a 210-bit long signature was obtained (hereafter referred as ChemAxon PF fp), which was as well generated for all 15 data sets used in this study.

**Table 2**. Characteristics of fingerprints, with the abbreviations used in this work.

| **Fingerprint** | **Abbreviation** | **Hashed** | **Length** |
|---|---|---|---|
| Estate fingerprint[42] | Est | No | 79 |
| MACCS fingerprint[43] | MACCS | No | 166 |
| PubChem fingerprint[41] | PubChem | No | 881 |
| Substructure fingerprint[41] | Substr | No | 308 |
| Klekotha-Roth fingerprint[44] | KRFP | No | 4860 |
| CDK fingerprint[45] | FP | Yes | 1024 |
| CDK extended fingerprint[41] | Ext | Yes | 1024 |
| CDK GraphOnly[41] | GraphOnly | Yes | 1024 |
| ChemAxon PF fingerprint[33] | ChemAxon PF fp | No | 210 |

Originally, Pharmacoprint was generated from the 2D structures without defined hydrogens. Nevertheless, in the present study, this approach was extended by adding hydrogens and by generating 3D representations. Finally, experiments were conducted for four representations of input structures (2D/3D with and without defined hydrogens). Both hydrogen addition and 3D conformation generation were performed using RDKit libraries. The generation of 3D structures and hydrogen addition do not change topological distances between the features, but they allow to truncate incorrectly drawn structures that may influence the model quality.

### *ML and reduction algorithms*

Pharmacoprint characterizes 39973 different bits under the settings described in the Methods section; for comparison, the 6 distance bins generate 25776 bits, while the 8 and 9 distance bins generate 58128 and 80604 bit-long vectors, respectively. In this light, it is impossible that thousands of chemical patterns occur in low-molecular-weight chemical compounds, and therefore, many on bits (i.e., equal to '1') are limited, which has two major consequences. On the one hand, ML algorithms are sensitive to noise that contaminate high-dimensional representations, while on the other hand, high-dimensional data generate additional computation cost of its processing.

A typical way of obtaining low-dimensional representation is to project data onto the most informative directions, which can be practically realized by the PCA (principal component analysis) algorithm.[46,47] To determine the number of resultant dimensions, two approaches were considered. In the first variant, a fixed-length representation was created by selecting 100 principal components. In the second variant, a final representation uses as many principal components as possible to describe 90% of the entire variance. The second approach makes it difficult to control the length of the resulting dimensions, but allows it to fit better to a particular dataset. Recent approaches for dimensionality reduction are based on the application of deep autoencoders (AEs).[48] AEs focus on finding a low-dimensional space that allows them to reconstruct the input signal. While PCA is restricted to orthogonal projections, AE can use arbitrary nonlinear mappings parametrized by neural networks. The proposed encoder network reduces input data to 100 latent dimensions by using 5 dense layers, and the decoder network uses symmetric layers. The choice of 100 dimensions was made in the analogy to one of the PCA variants considered.

From a computational perspective, PCA can be seen as a single-layer AE. Thus theoretically, the execution time of AE with 5 layers should be 5 times higher than the runtime of PCA. However, the exact processing time depends heavily on the implementation and the hardware used for the evaluation. Current neural networks were run on GPU, which significantly increases their speed.

Various ML methods were considered in the classification experiments. The first group included linear models: logistic regression (LR), linear support vector machines (LSVM), and support vector machines with RBF kernel (SVM). To take advantage of dimensionality reduction algorithms, both PCA and AE were combined with the aforementioned classification models.

The reduction algorithms considered here build a low-dimensional representation based on the internal characteristics of the data, ignoring completely the available class labels. To take this information into account, a supervised version of AE (sAE) was additionally studied.[49] In this approach, AE is trained jointly with a neural network classifier, where the classifier uses latent AE representation. The loss function of this model combines the reconstruction error with the classification cross-entropy. In contrast to typical AE, the representation generated by sAE should be better in discriminating active compounds from inactive ones. Two variants of classification models are considered for sAE: a single-layer linear classifier (hereafter called as SoftMax) and a neural network classifier built on 4 hidden layers (simply termed as NN).

In contrast to binary fingerprints which directly encodes subsequent chemical features, the representations produced by PCA and AE are continuous. Moreover, they are linear (in the case of PCA) or nonlinear (in the case of AE) combinations of all fingerprint attributes. Thus they cannot be easily interpreted as for fingerprints. The crucial advantage of sAE over the other two reduction algorithms is that its representation contains features that are needed to discriminate actives from inactives.

The implementations of PCA, SVM, and logistic regression were taken from sklearn Python library.[50] Neural networks were implemented in Pytorch toolkit.[51]

***Calculations and performance measures***

All the ML experiments were performed using a 10-fold cross-validation procedure. Matthews Correlation Coefficient (MCC, range: –1 to 1) was used as a measure of the quality of discrimination test in the paper, however, results for all conducted experiments along with additional statistics (Accuracy, balanced accuracy, recall, and area under the receiver operating characteristic curve (AUC ROC)) can be found in Table S1.[52]

$$MCC=\frac{TP\times TN-FP\times FN}{\sqrt{(TP+FP)(TP+FN)(TN+FP)(TN+FN)}}$$

$$Accuracy=\frac{TP+TN}{TP+TN+FP+FN}$$

$$\mathrm{Recal}\,l=\frac{TP}{TP+FN}$$

$$Bal_{acc}=\frac{\frac{TP}{TP+FN}+\frac{TN}{TN+FP}}{2}$$

TP is the number of true positives (actives labeled as actives), TN is the number of true negatives (inactives labeled as inactives), FP is the number of false positives (inactives labeled as actives), and FN is the number of false negatives (actives labeled as inactives). For MCC a maximal value of 1 indicates perfect agreement, 0 implies random classification, and the minimal value of –1 denotes inverse class assignment. Other parameters are in a range from 0 to 1.

Tanimoto coefficient (Tc) is similarity metric of two equal length bit strings calculated as follows:

$$Tc=\frac{c}{a+b-c}$$

where a is the number of positive bits for bit string A, b is the number of positive bits for string B, c is the number of common positive bits between strings A and B.

Normalized Mutual Information (NMI) measures the similarity between the clustering $C=\{C_1,C_2,...,C_n\}$ and the reference (true) grouping $R=\{R_1,R_2,...,R_m\}$ of the same dataset X. The index is obtained by normalizing the mutual information I(R,C), by the average of entropies (0.5(H(C)+H(R)).

$$NMI(R,C)=\frac{2I(R,C)}{H(R)+H(C)}=-\frac{2\sum_{i=1}^{n}\sum_{j=1}^{m}P(C_i\cap R_j)\log_2\frac{P(C_i\cap R_j)}{P(C_i)P(R_j)}}{\sum_{i=1}^{n}P(C_i)\log_2 P(C_i)+\sum_{j=1}^{m}P(R_j)\log_2 P(R_j)}$$

where P(A) denotes the probability that an element belongs to set A. NMI maximal value of 1 is attained for two identical partitions, while random clustering returns value 0.

## Results and Discussion

### *Comparison of Fingerprints*

A comparison of efficiency of different molecular fingerprints and Pharmacoprint was performed. Two types of an applied set of inactives (i.e., true inactives from ChEMBL and assumed inactives from ZINC) were used. Pharmacoprint was generated with the following settings: no reduction algorithm (all 39973 bits were used, the longest fingerprint generated in PaDEL-Descriptor software (Klekotha-Roth fingerprint) was 8-fold shorter (4860 bits)), and 3D representation with defined positions of hydrogens was generated. All calculations were performed using three ML methods (SVM, LSVM, and LR).

The results (Figure 3) showed that Pharmacoprint outperformed all tested fingerprints in the separation of true actives from true inactives. Pharmacoprint obtained the highest median MCC and average MCC for all experiments. The mean MCC of Pharmacoprint was 12 percentage points higher than that of the second best-performing fingerprint (Extended with an average MCC=0.635). Discrimination of the actives from the ZINC compounds is a much less demanding task,[40] and five out of nine tested fingerprints achieved median MCC above 0.9. In this case, the performance of Pharmacoprint was comparable to that of the widely used KRFP (median MCC for KFRP and Pharmacoprint was 0.980)[38]. Other statistics are shown in SI (Figures S4–S7). Although the results are ML method-dependent, only in SVM experiments, Ext fingerprint could compete with Pharmacoprint (Figures S1–S3).

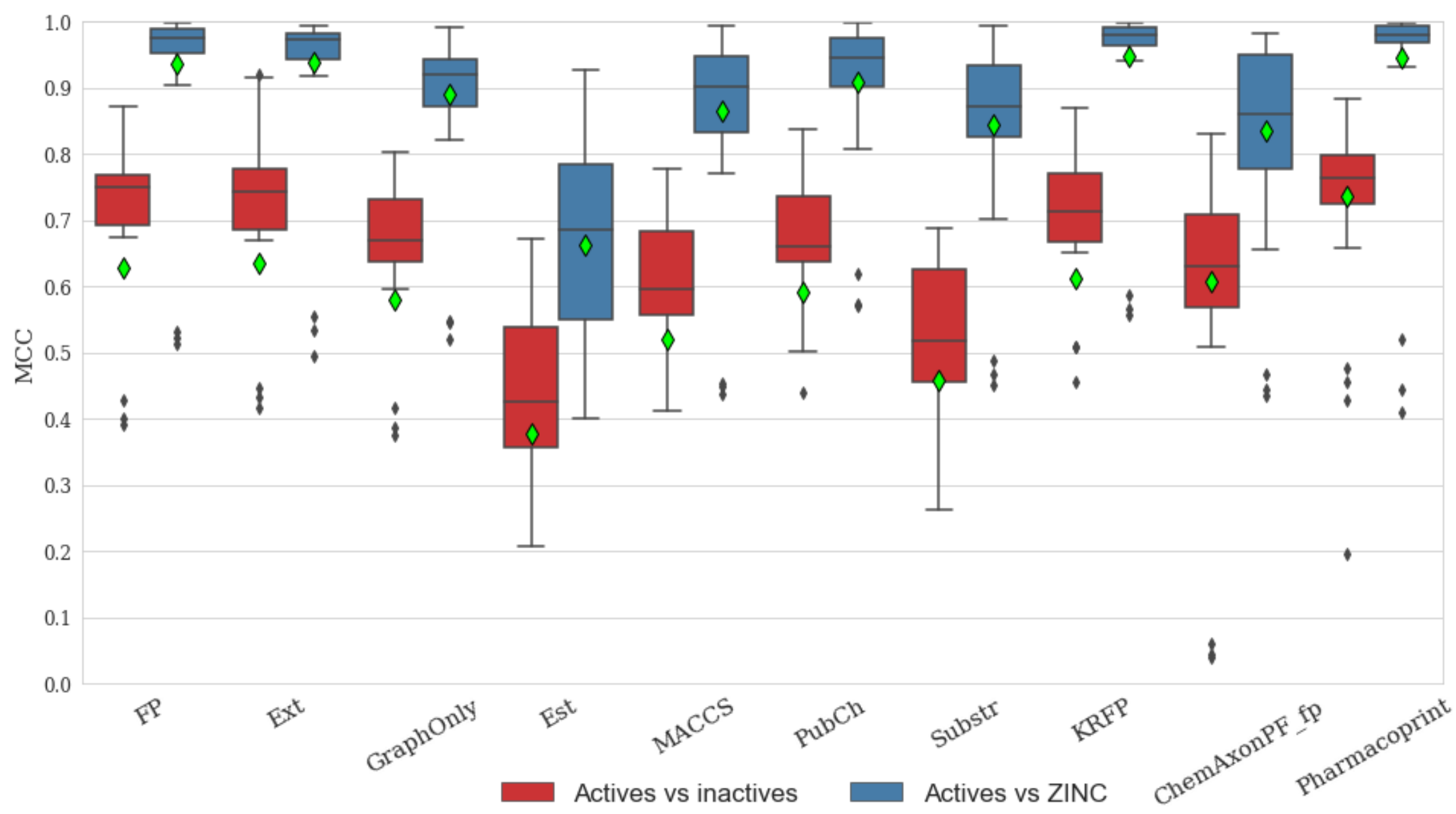

**Figure 3**. Boxplot of MCC values (calculated for three ML methods and 15 different targets) for all fingerprints analyzed in this study. Quality of actives/inactives discrimination is significantly different when active compounds were mixed with compounds with confirmed inactivity and compounds with putative inactivity from ZINC. Median values are marked as a horizontal line within the box, average values are denoted as green diamonds, and outliers are denoted as small black diamonds (outliers with MCC below 0 are not shown).

Pharmacoprint was compared with ChemAxon PF fp (separately for targets and ML approaches). Results (Table 3) showed the superiority of Pharmacoprint which achieved higher MCC values in 37 cases out of 45. Nevertheless, ChemAxon PF fp can be applied as input data for SVM experiments or for sets with a balanced but limited amount of data (like NMDA). The comparison of ChemAxon PF fp with Pharmacoprint for other statistical parameters is presented in Table S2.

**Table 3**. A comparison of the difference of MCC values (calculated for three ML methods and 15 different targets) for Pharamacoprint and ChemAxon PF fp. A positive value indicates that the Pharmacoprint was better.

| **Receptor** | **LSVM** | **LR** | **SVM** |
|---|---|---|---|
| 5HT2A | 0.154 | 0.150 | –0.001 |
| 5HT2C | 0.125 | 0.185 | –0.008 |
| 5HT6 | 0.063 | 0.054 | 0.010 |
| catB | 0.138 | 0.175 | 0.042 |
| catL | 0.153 | 0.224 | 0.091 |
| D2 | 0.224 | 0.241 | 0.036 |
| delta | 0.196 | 0.213 | 0.031 |
| HIVint | 0.195 | 0.161 | 0.097 |
| HIVprot | 0.099 | 0.130 | 0.013 |
| HIVrev | –0.021 | 0.074 | –0.039 |
| kappa | 0.182 | 0.200 | 0.070 |
| mi | 0.185 | 0.211 | 0.098 |
| NMDA | –0.109 | –0.006 | –0.103 |
| NOP | 0.061 | 0.070 | –0.039 |
| NPC1 | 0.117 | 0.121 | 0.193 |

***Impact of the input molecule representation***

The subsequent experiments were conducted only for Pharmacoprint representation with no reduction algorithm to elucidate whether chemical structures should be sanitized before fingerprint generation.

The results (Figure 4) confirmed that proper preparation of chemical structures before Pharmacoprint generation is crucial for obtaining high-quality results. 3D optimized structures (with the use of RDKit libraries) returned better results, regardless of the ML method applied. Defining the positions of hydrogens also led to further improvement of results by skipping uncertain data (the only exception noted was when linear SVM was applied); therefore, for ensuring the highest quality of results, compounds should be prepared as 3D with defined positions of hydrogens. The application of this procedure allows both for neglecting incorrectly written structures and for eliminating extremely large structures (with molecular weight > 1000Da) from the training set. Large molecules are especially "dangerous" for developing ML models on the representation with a limited number of bits because they possess many pharmacophore features that greatly increase the amount of '1' in bit string representation. In Pharmacoprint, when the amount of pharmacophore features increases two times, the number of on bits is 10-fold higher. Therefore, such molecules introduce much noise. The highest increase in the quality of results is observed when compounds are converted from 2D to 3D, which ensures the elimination of high-molecular-weight compounds. If Pharmacoprint is generated for an already sanitized and checked set of compounds, fingerprint representation and results are the same. Therefore, automatic truncation is a mandatory step for high-quality results. Nevertheless, it is worth noting that Pharmacoprint as a 2D descriptor (even derived from the 3D structure) cannot reflect the stereochemistry which can be achieved by developing a classical 3D pharmacophore model.

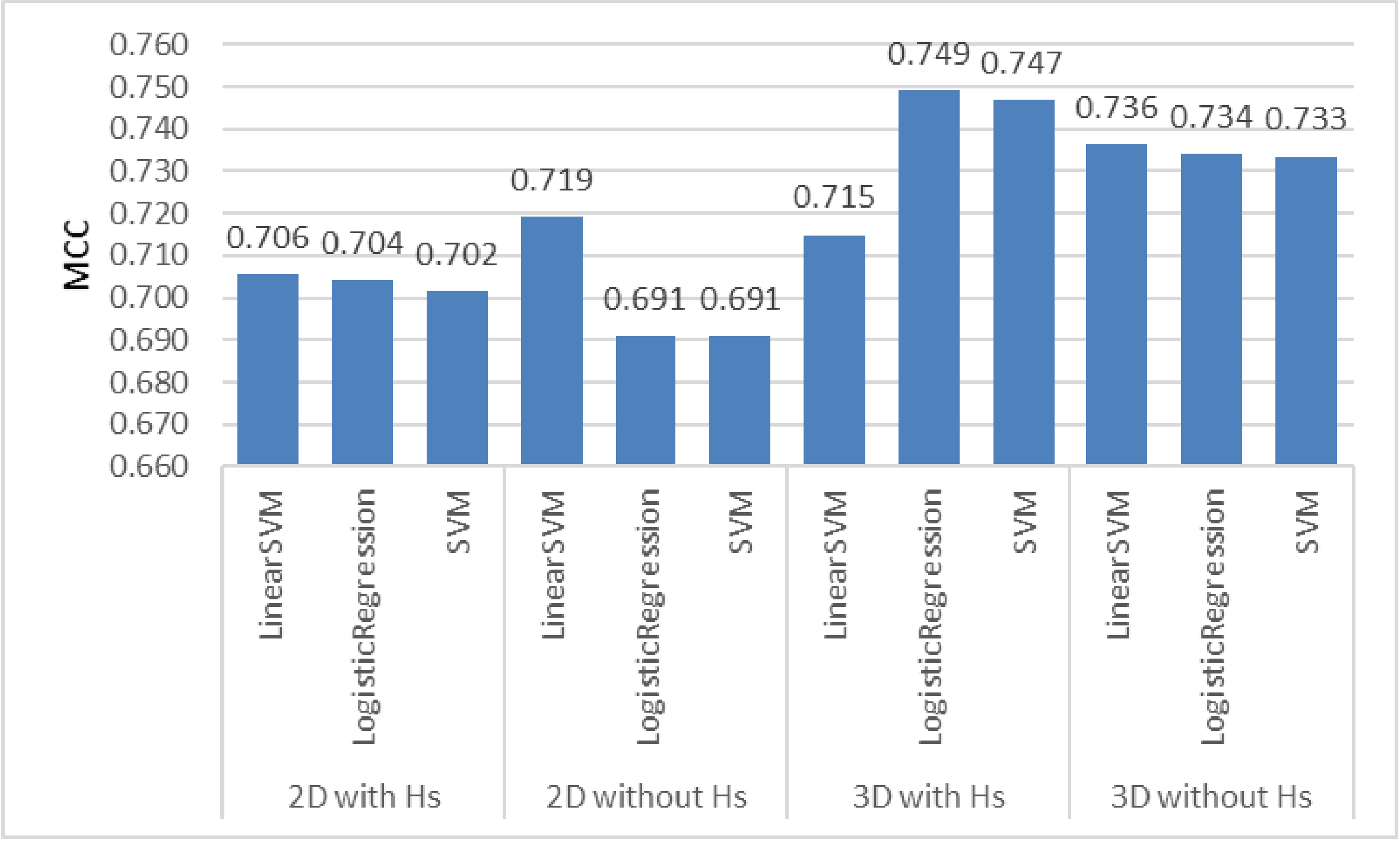

**Figure 4**. Average MCC (calculated for 15 different biological targets) values for different representations of the input molecules obtained in classification experiments for three different machine learning approaches. Better separation of actives from true inactives is observed when Pharmacoprint is generated from a 3D structure with defined hydrogens.

***Evaluation of different ML methods and fingerprint reduction algorithms***

The use of high-resolution molecular fingerprints (such as KRFP) often results in high time consumption. Previous studies have shown that the application of the reduction algorithm allows not only to save CPU time but also to improve the results.[38,53,54] Pharmacoprint is an extremely long bit string, where the frequency of occurrence of '1' instead of '0' is limited, which creates an excellent opportunity for the application of reduction algorithms. For this purpose, 100-dimensional representations generated by PCA, AE, and sAE were considered. An additional case which considers the number of principal components that explain 90% of the variance was also studied. While PCA and AE could be combined with any classifier, sAE can only be used with neural network classifiers because the entire system is trained jointly.

Although reduction of the fingerprint to the 100 principal components returned the worst quality of classification (Figure 5, average MCC for LR and LSVM classifiers was between 0.6 and 0.7, other statistics are shown in Figures S8–S11), neglecting such reduction approaches did not yield worse results when the reduction in time consumption was considered important for the user. The use of more complex nonlinear SVM increased the average MCC to 0.687. Adapting the number of principal components to the particular dataset (PCA 90%) yielded better results; this was not surprising because more than 100 features were used to explain 90% of the total variance. Application of the autoencoder with linear classifiers (LSVM and LR) further reduced the average MCC to below 0.5, but combining the autoencoder with a more sophisticated neural network classifier increased the average MMC to above 0.750. This analysis showed that nonlinear classifiers (SVM for PCA and NN for AE) can obtain fairly good results using low-quality representations. The results obtained for the neural network were further improved up to 0.774 when a supervised autoencoder was used instead of an unsupervised one. A very simple SoftMax classifier combined with sAE yielded only slightly worse performance, which confirms that sAE builds representations that are both compact and discriminative. The neural network was found to be the most robust ML method and yielded even better results than that obtained using the original Pharmacoprint when no reduction algorithm is applied (the best value of the average MCC was 0.752 for LSVM).

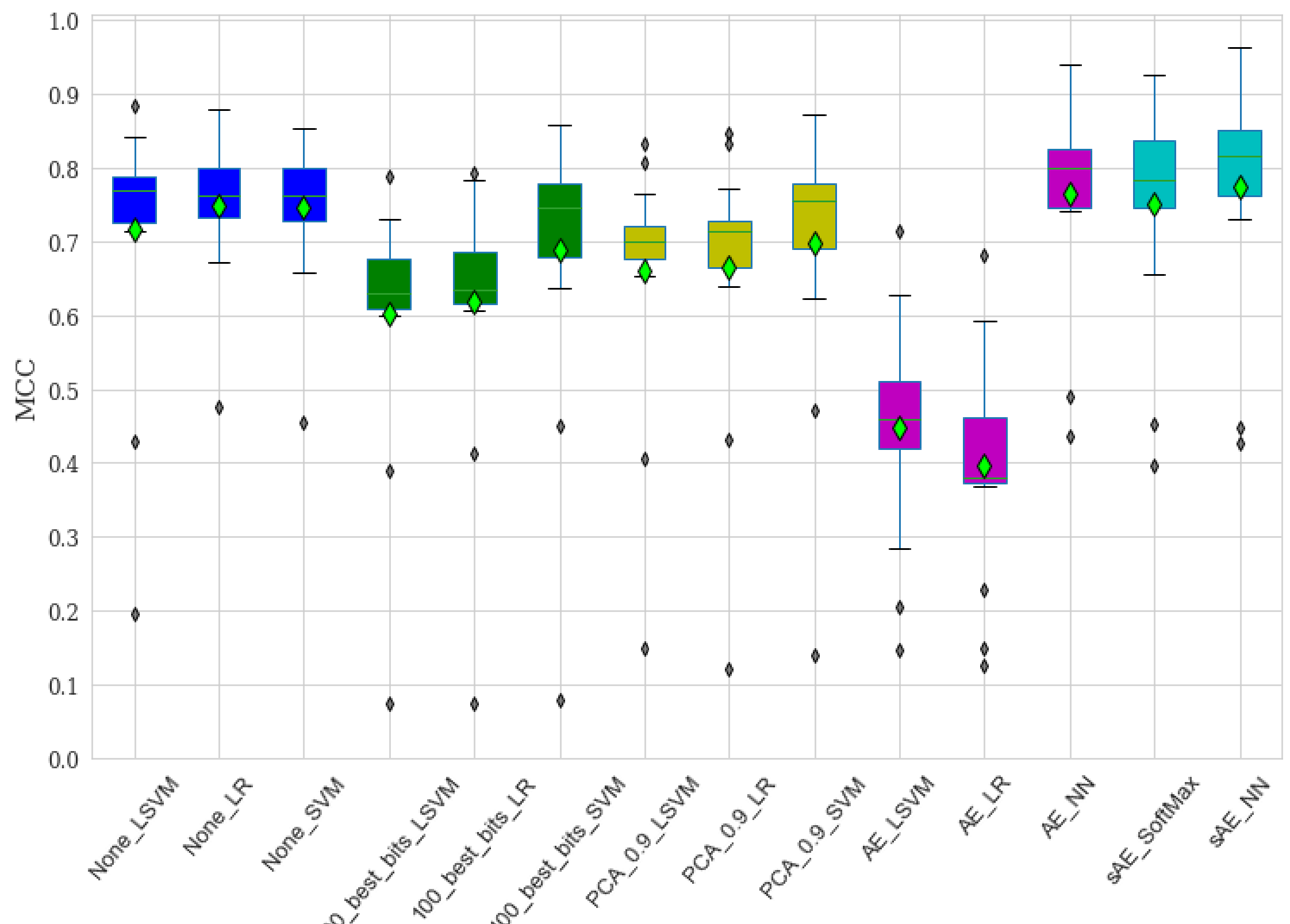

**Figure 5**. Boxplot of MCC values (calculated for 15 different biological targets, only for Pharmacoprint) for different fingerprint reduction approaches obtained in classification experiments (separation of actives from true inactives) for different machine learning methods. The best results were obtained when a supervised autoencoder was applied along with the neural network. Not all combinations of machine learning methods and reduction types were tested. Median values are marked as a horizontal line within the box, average values are denoted as green diamonds, and outliers are denoted as small black diamonds (outliers with MCC below 0 not shown). Boxes are grouped according to the applied reduction algorithm.

***Target dependency***

When analyzing the results for various targets (Figure 6, other statistics are shown in Figures S12–S15), it was observed that the results for some targets outperformed other results regardless of which ML/reduction algorithm combination was applied. Especially, good results were observed for biological targets that have the highest number of active compounds (e.g., delta opioid receptor or serotonin 5-$HT_6$ receptor). The worst results were obtained for NMDA and NPC1 receptors, which have a relatively low number of active compounds. The poor results obtained for these sets are shown in Figures 3 and 5 where they are always outliers. However, a similar relationship was not observed for HIV integrase inhibitors, which, despite having a similar number of active compounds, allowed to create very efficient classification models (the MCC value reached up to 0.962, which is the maximum value achieved in the described study). This is because HIV integrase inhibitors and inactives explore broader chemical space than NMDA receptor ligands and inactives (Table S3). A higher inner dissimilarity was observed among NPC1 actives and inactives than for HIV integrase inhibitors, but in this case, poor results are due to misbalance between the number of actives and inactive (~1:161). The results confirmed that the best method of fingerprint reduction is by using the supervised autoencoder; for most of the targets, the highest MCC values were obtained for this method, which is particularly efficient when combined with a neural network.

| | AE_LSVM | AE_LR | AE_NN | LSVM | Logistic Regression | sAE_SoftMax | sAE_NN | SVM |
|---|---|---|---|---|---|---|---|---|
| 5-HT$_{2A}$ | 0.482 | 0.446 | 0.805 | 0.714 | 0.714 | 0.783 | 0.820 | 0.680 |
| 5-HT$_{2C}$ | 0.411 | 0.381 | 0.791 | 0.720 | 0.733 | 0.764 | 0.792 | 0.721 |
| 5-HT$_{6}$ | 0.430 | 0.380 | 0.834 | 0.728 | 0.672 | 0.855 | 0.861 | 0.658 |
| catB | 0.471 | 0.379 | 0.741 | 0.780 | 0.764 | 0.656 | 0.730 | 0.789 |
| catL | 0.204 | 0.149 | 0.800 | 0.781 | 0.776 | 0.797 | 0.816 | 0.801 |
| D$_{2}$ | 0.479 | 0.413 | 0.761 | 0.770 | 0.759 | 0.748 | 0.782 | 0.763 |
| delta | 0.712 | 0.680 | 0.886 | 0.843 | 0.849 | 0.880 | 0.898 | 0.852 |
| HIVint | 0.444 | 0.369 | 0.940 | 0.883 | 0.880 | 0.925 | 0.962 | 0.849 |
| HIVprot | 0.541 | 0.478 | 0.815 | 0.752 | 0.733 | 0.836 | 0.841 | 0.761 |
| HIVrev | 0.434 | 0.378 | 0.750 | 0.740 | 0.736 | 0.779 | 0.787 | 0.750 |
| kappa | 0.592 | 0.542 | 0.855 | 0.770 | 0.767 | 0.821 | 0.874 | 0.762 |
| mi | 0.626 | 0.591 | 0.818 | 0.828 | 0.823 | 0.838 | 0.830 | 0.821 |
| NMDA | 0.285 | 0.230 | 0.489 | 0.429 | 0.476 | 0.452 | 0.447 | 0.455 |
| NOP | 0.458 | 0.390 | 0.741 | 0.793 | 0.807 | 0.744 | 0.742 | 0.795 |
| NPC1 | 0.148 | 0.125 | 0.436 | 0.195 | 0.426 | 0.397 | 0.501 | 0.465 |

**Figure 6**. MCC values were obtained for different biological targets with different machine learning methods and reduction algorithms for Pharmacoprint. A reduction algorithm is indicated before the name of the ML method, and AE and sAE denote autoencoder and supervised autoencoder, respectively. No underscore implies that no reduction algorithm was applied.

### *Clustering of chemical libraries*

As an example of chemical libraries, we considered a dataset of 3616 compounds acting on the 5-HT$_{1A}$ receptor (with affinity expressed as $K_i$ or equivalent below 100 nM). In our previous study, this set was manually divided into 28 structural subgroups based on expert knowledge, was used reference clustering.[22,53] Such manual procedure is time and effort consuming thus application of this methodology for other compound sets would be impossible. Therefore, all the fingerprints described in this paper were used for the automatic, hierarchical clustering using the Tanimoto metric and complete linkage function.[53] Next, generated clusters were compared with manual grouping. The results (Table 4.) indicate that clustering with Pharmacoprint returned results the most similar to the reference clustering and should consider for automatic clustering purposes in the future.

**Table 4.** NMI values between manual reference clustering and clustering are performed with the application of fingerprints described within this paper.

| **Fingerprint** | **NMI** |
| --- | --- |
| KRFP | 0.356 |
| MACCS | 0.297 |
| Pharmacoprint | 0.376 |
| Est | 0.268 |
| GraphOnly | 0.275 |
| FP | 0.352 |
| Ext | 0.361 |
| Substr | 0.270 |
| PubChem | 0.283 |
| ChemAxon PF fp | 0.295 |

***Similarity searching***

Another application of the Pharmacoprint concerns similarity searching. Among 3616 5-$HT_{1A}$ receptor ligands, the most active compound (ChEMBL42393, $K_i$ = 0.28nM) was selected as the reference and then compared with the rest of the compounds using all fingerprints discussed in this study along with the Tanimoto similarity metric. Results (Figures 7 and S16) showed that Pharmacoprint can be used in similarity searching with as good effect as other fingerprints because the structures identified as the most similar differ only in minor details of the chemical structure from the reference compound. In addition, unlike Est, MACCS and ChemAxon PF fp, none of the structures achieved a Tanimoto coefficient equal to 1, which indicates that Pharmacoprint can also be used to remove duplicates from the databases.

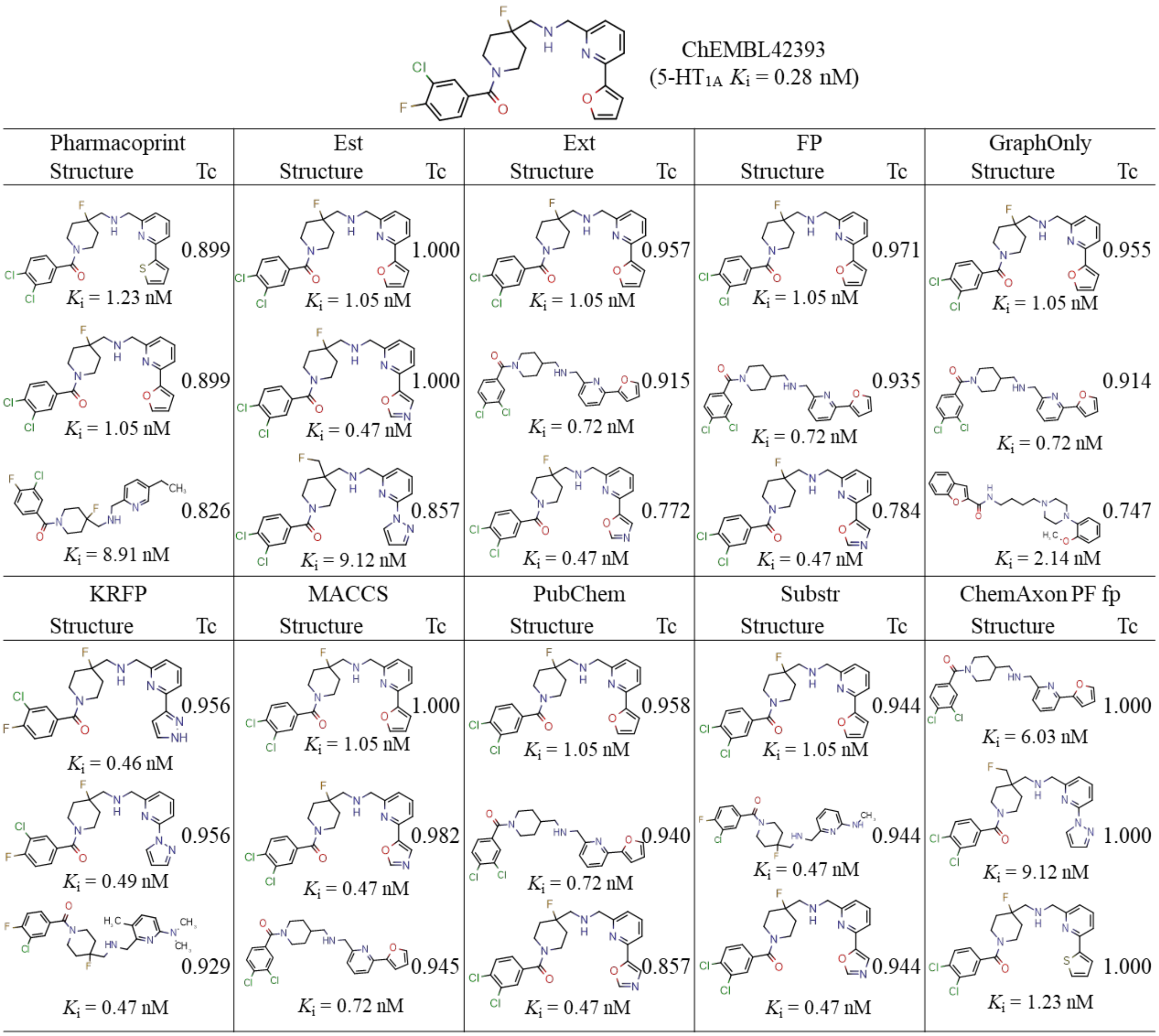

**Figure 7.** Structure of the ChEMBL42393 – the most active compounds among all 5-$HT_{1A}$R actives analyzed within this study along with three most similar compounds (with a value of the Tanimoto metric) per fingerprint type used for the similarity searching.

## Conclusions

This paper describes a fingerprint characterizing the presence of pharmacophore features in the structure of chemical compounds. Pharmacoprint is one of the longest fingerprint; the full signature generated under the settings used in this paper consists of 39973 bits, which, however, is susceptible to reduction.

A multilevel analysis of the results obtained in this study allowed to define the most robust conditions of Pharmacoprint generation and application. This representation when used as an input to the ML method outperformed eight other popular molecular fingerprints in separating true actives from inactives. Further optimization allowed to define the optimal form of molecule used for fingerprint generation, ML method, and an algorithm for fingerprint reduction.

Sanitization of the input molecules as 3D with a defined position of hydrogen atoms allowed to obtain the highest performance in the ML experiments. The full-length fingerprint could efficiently separate actives from inactives; however, the use of some reduction algorithms such as supervised encoder not only reduced time consumption but also improved statistical parameters, especially when a neural network was applied.

In summary, if Pharmacoprint was generated from a 3D input structure with defined hydrogen atoms and applied as training data for a neural network with supervised autoencoder function, an MCC value of up to 0.962 could be achieved in discrimination tests.

In addition to the classification experiments, it was demonstrated that Pharmacoprint can be successfully applied in other areas of cheminformatics. It showed the best performance in reproducing reference clustering among all the fingerprints analyzed within this study. Pharmacoprint can be also successfully applied in similarity searching.

## Data and software availability

The code and datasets used in this study are available at https://github.com/lstruski/Pharmacoprint. All the instructions can be found in README file.

## Corresponding Author

*Rafał Kurczab - Maj Institute of Pharmacology Polish Academy of Sciences, Smetna 12 Street, 31-343 Cracow, Poland; orcid.org/0000-0002-9555-3905; Email: kurczab@if-pan.krakow.pl

## Authors' contributions

Dawid Warszycki and Rafał Kurczab conceived and supervised the study. Dawid Warszycki, Marek Śmieja, and Rafał Kurczab designed the study plan. Marek Śmieja, Łukasz Struski, and Rafał Kafel were responsible for implementing the algorithms. Dawid Warszycki and Rafał Kafel undertook the chemical descriptors generation and bioassay data curation. Marek Śmieja and Łukasz Struski provided the machine learning implementation. Dawid Warszycki and Marek Śmieja drafted the manuscript. All authors revised the manuscript. All authors read and approved the final manuscript.

## Funding

The works of the authors Dawid Warszycki, Marek Śmieja, Rafał Kafel, and Rafał Kurczab were supported by the Polish National Centre for Research and Development (Grant LIDER/37/0137/L-9/17/NCBR/2018). The work of Łukasz Struski was funded by the National

Science Centre, Poland (research grant no. 2020/39/D/ST6/01332). For the purpose of Open Access, the author has applied a CC-BY public copyright licence to any Author Accepted Manuscript (AAM) version arising from this submission.

**Note**

The authors declare no competing interests.

**Abbreviations**

AE, autoencoder; AUROC, area under the receiver operating characteristic curve; LSVM, linear support vector machine; FN, false negative; FP, false positive; LR, logistic regression; MCC, Matthews correlation coefficient; ML, machine learning; NMI, normalized mutual information; NN, neural network; PCA, principal component analysis; PF, pharmacophore features; sAE, supervised autoencoder; SVM, support vector machine; Tc, Tanimoto coefficient; TN, true negatives; TP, true positives

## References

(1) Czarnecki, W. M.; Tabor, J. Multithreshold Entropy Linear Classifier: Theory and Applications. *Expert Syst. Appl.* **2015**, *42* (13), 5591–5606. https://doi.org/10.1016/j.eswa.2015.03.007.

(2) Staroń, J.; Warszycki, D.; Kalinowska-Tłuścik, J.; Satała, G.; Bojarski, A. J. Rational Design of 5-HT 6 R Ligands Using a Bioisosteric Strategy: Synthesis, Biological Evaluation and Molecular Modelling. *RSC Adv.* **2015**, *5* (33), 25806–25815. https://doi.org/10.1039/C5RA00054H.

(3) Smusz, S.; Mordalski, S.; Witek, J.; Rataj, K.; Kafel, R.; Bojarski, A. J. Multi-Step Protocol for Automatic Evaluation of Docking Results Based on Machine Learning Methods--A Case Study of Serotonin Receptors 5-HT(6) and 5-HT(7). *J. Chem. Inf. Model.* **2015**, *55* (4), 823–832. https://doi.org/10.1021/ci500564b.

(4) Smusz, S.; Kurczab, R.; Satała, G.; Bojarski, A. J. Fingerprint-Based Consensus Virtual Screening towards Structurally New 5-HT6R Ligands. *Bioorg. Med. Chem. Lett.* **2015**, *25* (9), 1827–1830. https://doi.org/10.1016/j.bmcl.2015.03.049.

(5) Witek, J.; Smusz, S.; Rataj, K.; Mordalski, S.; Bojarski, A. J. An Application of Machine Learning Methods to Structural Interaction Fingerprints—a Case Study of Kinase Inhibitors. *Bioorg. Med. Chem. Lett.* **2014**, *24* (2), 580–585. https://doi.org/10.1016/j.bmcl.2013.12.017.

(6) Gabrielsen, M.; Kurczab, R.; Ravna, A. W.; Kufareva, I.; Abagyan, R.; Chilmonczyk, Z.; Bojarski, A. J.; Sylte, I. Molecular Mechanism of Serotonin Transporter Inhibition Elucidated by a New Flexible Docking Protocol. *Eur. J. Med. Chem.* **2012**, *47*, 24–37. https://doi.org/10.1016/j.ejmech.2011.09.056.

(7) Zajdel, P.; Kurczab, R.; Grychowska, K.; Satała, G.; Pawłowski, M.; Bojarski, A. J. The Multiobjective Based Design, Synthesis and Evaluation of the Arylsulfonamide/Amide Derivatives of Aryloxyethyl- and Arylthioethyl- Piperidines and Pyrrolidines as a Novel Class of Potent 5-HT(7) Receptor Antagonists. *Eur. J. Med. Chem.* **2012**, *56*, 348–360.

https://doi.org/10.1016/j.ejmech.2012.07.043.

(8) Kurczab, R.; Nowak, M.; Chilmonczyk, Z.; Sylte, I.; Bojarski, A. J. The Development and Validation of a Novel Virtual Screening Cascade Protocol to Identify Potential Serotonin 5-HT(7)R Antagonists. *Bioorg. Med. Chem. Lett.* **2010**, *20* (8), 2465–2468. https://doi.org/10.1016/j.bmcl.2010.03.012.

(9) Kurczyk, A.; Warszycki, D.; Musiol, R.; Kafel, R.; Bojarski, A. J.; Polanski, J. Ligand-Based Virtual Screening in a Search for Novel Anti-HIV-1 Chemotypes. *J. Chem. Inf. Model.* **2015**, *55* (10), 2168–2177. https://doi.org/10.1021/acs.jcim.5b00295.

(10) Warszycki, D.; Rueda, M.; Mordalski, S.; Kristiansen, K.; Satała, G.; Rataj, K.; Chilmonczyk, Z.; Sylte, I.; Abagyan, R.; Bojarski, A. J. From Homology Models to a Set of Predictive Binding Pockets–a 5-HT 1A Receptor Case Study. *J. Chem. Inf. Model.* **2017**, *57* (2), 311–321. https://doi.org/10.1021/acs.jcim.6b00263.

(11) Śmieja, M.; Tabor, J.; Spurek, P. SVM with a Neutral Class. *Pattern Anal. Appl.* **2019**, *22* (2), 573–582. https://doi.org/10.1007/s10044-017-0654-3.

(12) Śmieja, M.; Hajto, K.; Tabor, J. Efficient Mixture Model for Clustering of Sparse High Dimensional Binary Data. *Data Min. Knowl. Discov.* **2019**, *33* (6), 1583–1624. https://doi.org/10.1007/s10618-019-00635-1.

(13) Śmieja, M.; Geiger, B. C. Semi-Supervised Cross-Entropy Clustering with Information Bottleneck Constraint. *Inf. Sci. (Ny).* **2017**, *421*, 254–271. https://doi.org/10.1016/j.ins.2017.07.016.

(14) Podlewska, S.; Kafel, R. MetStabOn—Online Platform for Metabolic Stability Predictions. *Int. J. Mol. Sci.* **2018**, *19* (4), 1040. https://doi.org/10.3390/ijms19041040.

(15) Rataj, K.; Czarnecki, W.; Podlewska, S.; Pocha, A.; Bojarski, A. Substructural Connectivity Fingerprint and Extreme Entropy Machines—A New Method of Compound Representation and Analysis. *Molecules* **2018**, *23* (6), 1242. https://doi.org/10.3390/molecules23061242.

(16) Czarnecki, W.; Podlewska, S.; Bojarski, A. Extremely Randomized Machine Learning Methods for Compound Activity Prediction. *Molecules* **2015**, *20* (11), 20107–20117. https://doi.org/10.3390/molecules201119679.

(17) Czarnecki, W. M.; Podlewska, S.; Bojarski, A. J. Robust Optimization of SVM Hyperparameters in the Classification of Bioactive Compounds. *J. Cheminform.* **2015**, *7* (1), 38. https://doi.org/10.1186/s13321-015-0088-0.

(18) Kurczab, R.; Bojarski, A. J. The Influence of the Negative-Positive Ratio and Screening Database Size on the Performance of Machine Learning-Based Virtual Screening. *PLoS One* **2017**, *12* (4), e0175410. https://doi.org/10.1371/journal.pone.0175410.

(19) IUPAC, Glossary of Terms Used in Medicinal Chemistry, Http:// Www.Chem.Qmul.Ac.Uk/Iupac/Medchem.

(20) Freyd, T.; Warszycki, D.; Mordalski, S.; Bojarski, A. J.; Sylte, I.; Gabrielsen, M. Ligand-Guided Homology Modelling of the GABAB2 Subunit of the GABAB Receptor. *PLoS One* **2017**, *12* (3), e0173889. https://doi.org/10.1371/journal.pone.0173889.

(21) Sanders, M. P. a; Verhoeven, S.; de Graaf, C.; Roumen, L.; Vroling, B.; Nabuurs, S. B.; de Vlieg, J.; Klomp, J. P. G. Snooker: A Structure-Based Pharmacophore Generation Tool Applied to Class A GPCRs. *J. Chem. Inf. Model.* **2011**, *51* (9), 2277–2292. https://doi.org/10.1021/ci200088d.

(22) Warszycki, D.; Mordalski, S.; Kristiansen, K.; Kafel, R.; Sylte, I.; Chilmonczyk, Z.; Bojarski, A. J. A Linear Combination of Pharmacophore Hypotheses as a New Tool in Search of New Active Compounds--an Application for 5-HT1A Receptor Ligands. *PLoS One* **2013**, *8* (12), e84510. https://doi.org/10.1371/journal.pone.0084510.

(23) Svensson, F.; Karlén, A.; Sköld, C. Virtual Screening Data Fusion Using Both Structure- and Ligand-Based Methods. *J. Chem. Inf. Model.* **2012**, *52* (1), 225–232. https://doi.org/10.1021/ci2004835.

(24) Durdagi, S.; Duff, H. J.; Noskov, S. Y. Combined Receptor and Ligand-Based Approach to the Universal Pharmacophore Model Development for Studies of Drug Blockade to the HERG1 Pore Domain. *J. Chem. Inf. Model.* **2011**, *51* (2), 463–474. https://doi.org/10.1021/ci100409y.

(25) Manepalli, S.; Geffert, L. M.; Surratt, C. K.; Madura, J. D. Discovery of Novel Selective Serotonin Reuptake Inhibitors through Development of a Protein-Based Pharmacophore. *J. Chem. Inf. Model.* **2011**, *51* (9), 2417–2426. https://doi.org/10.1021/ci200280m.

(26) Chiu, T.-L.; Amin, E. a. Development of a Comprehensive, Validated Pharmacophore Hypothesis for Anthrax Toxin Lethal Factor (LF) Inhibitors Using Genetic Algorithms, Pareto Scoring, and Structural Biology. *J. Chem. Inf. Model.* **2012**, *52* (7), 1886–1897. https://doi.org/10.1021/ci300121p.

(27) Zajdel, P.; Kurczab, R.; Grychowska, K.; Satała, G.; Pawłowski, M.; Bojarski, A. J. The Multiobjective Based Design, Synthesis and Evaluation of the Arylsulfonamide/Amide Derivatives of Aryloxyethyl- and Arylthioethyl- Piperidines and Pyrrolidines as a Novel Class of Potent 5-$HT_7$ Receptor Antagonists. *Eur. J. Med. Chem.* **2012**, *56*, 348–360. https://doi.org/10.1016/j.ejmech.2012.07.043.

(28) Bonachéra, F.; Parent, B.; Barbosa, F.; Froloff, N.; Horvath, D. Fuzzy Tricentric Pharmacophore Fingerprints. 1. Topological Fuzzy Pharmacophore Triplets and Adapted Molecular Similarity Scoring Schemes. *J. Chem. Inf. Model.* **2006**, *46* (6), 2457–2477. https://doi.org/10.1021/ci6002416.

(29) Wood, D. J.; Vlieg, J. de; Wagener, M.; Ritschel, T. Pharmacophore Fingerprint-Based Approach to Binding Site Subpocket Similarity and Its Application to Bioisostere Replacement. *J. Chem. Inf. Model.* **2012**, *52* (8), 2031–2043. https://doi.org/10.1021/ci3000776.

(30) Askjaer, S.; Langgård, M. Combining Pharmacophore Fingerprints and PLS-Discriminant

Analysis for Virtual Screening and SAR Elucidation. *J. Chem. Inf. Model.* **2008**, *48* (3), 476–488. https://doi.org/10.1021/ci700356w.

(31) McGregor, M. J.; Muskal, S. M. Pharmacophore Fingerprinting. 1. Application to QSAR and Focused Library Design. *J. Chem. Inf. Comput. Sci.* **1999**, *39* (3), 569–574. https://doi.org/10.1021/ci980159j.

(32) McGregor, M. J.; Muskal, S. M. Pharmacophore Fingerprinting. 2. Application to Primary Library Design. *J. Chem. Inf. Comput. Sci.* **2000**, *40* (1), 117–125.

(33) Marvin 20.15.24, 2020, ChemAxon (Http://Www.Chemaxon.Com).

(34) Wagener, M.; Lommerse, J. P. M. The Quest for Bioisosteric Replacements. *J. Chem. Inf. Model.* **2006**, *46* (2), 677–685. https://doi.org/10.1021/ci0503964.

(35) Gobbi, A.; Poppinger, D. Genetic Optimization of Combinatorial Libraries. *Biotechnol. Bioeng.* **1998**, *61* (1), 47–54. https://doi.org/10.1002/(SICI)1097-0290(199824)61:1<47::AID-BIT9>3.0.CO;2-Z.

(36) Gaulton, A.; Hersey, A.; Nowotka, M.; Bento, A. P.; Chambers, J.; Mendez, D.; Mutowo, P.; Atkinson, F.; Bellis, L. J.; Cibrián-Uhalte, E.; Davies, M.; Dedman, N.; Karlsson, A.; Magariños, M. P.; Overington, J. P.; Papadatos, G.; Smit, I.; Leach, A. R. The ChEMBL Database in 2017. *Nucleic Acids Res.* **2017**, *45* (D1), D945–D954. https://doi.org/10.1093/nar/gkw1074.

(37) Kalliokoski, T.; Kramer, C.; Vulpetti, A.; Gedeck, P. Comparability of Mixed IC$_{50}$ Data - a Statistical Analysis. *PLoS One* **2013**, *8* (4), e61007. https://doi.org/10.1371/journal.pone.0061007.

(38) Śmieja, M.; Warszycki, D. Average Information Content Maximization--A New Approach for Fingerprint Hybridization and Reduction. *PLoS One* **2016**, *11* (1), e0146666. https://doi.org/10.1371/journal.pone.0146666.

(39) Smusz, S.; Czarnecki, W. M.; Warszycki, D.; Bojarski, A. J. Exploiting Uncertainty Measures in Compounds Activity Prediction Using Support Vector Machines. *Bioorg. Med. Chem. Lett.* **2014**, *25* (1), 100–105. https://doi.org/10.1016/j.bmcl.2014.11.005.

(40) Smusz, S.; Kurczab, R.; Bojarski, A. J. The Influence of the Inactives Subset Generation on the Performance of Machine Learning Methods. *J. Cheminform.* **2013**, *5* (1), 17. https://doi.org/10.1186/1758-2946-5-17.

(41) Yap, C. W. PaDEL-Descriptor: An Open Source Software to Calculate Molecular Descriptors and Fingerprints. *J. Comput. Chem.* **2011**, *32* (7), 1466–1474. https://doi.org/10.1002/jcc.21707.

(42) Hall, L. H.; Kier, L. B. Electrotopological State Indices for Atom Types: A Novel Combination of Electronic, Topological, and Valence State Information. *J. Chem. Inf. Model.* **1995**, *35* (6), 1039–1045. https://doi.org/10.1021/ci00028a014.

(43) Ewing, T.; Baber, J. C.; Feher, M. Novel 2D Fingerprints for Ligand-Based Virtual

Screening. *J. Chem. Inf. Model.* **2006**, *46* (6), 2423–2431. https://doi.org/10.1021/ci060155b.

(44) Klekota, J.; Roth, F. P. Chemical Substructures That Enrich for Biological Activity. *Bioinformatics* **2008**, *24* (21), 2518–2525. https://doi.org/10.1093/bioinformatics/btn479.

(45) Steinbeck, C.; Han, Y.; Kuhn, S.; Horlacher, O.; Luttmann, E.; Willighagen, E. The Chemistry Development Kit (CDK): An Open-Source Java Library for Chemo- and Bioinformatics. *J. Chem. Inf. Comput. Sci.* **2003**, *43* (2), 493–500. https://doi.org/10.1021/ci025584y.

(46) Alpaydin, E. *Machine Learning: The New AI*; MIT Press, 2016.

(47) Alpaydin, E. *Introduction to Machine Learning.*; MIT Press, 2014.

(48) Wang, Y.; Yao, H.; Zhao, S. Auto-Encoder Based Dimensionality Reduction. *Neurocomputing* **2016**, *184*, 232–242. https://doi.org/10.1016/j.neucom.2015.08.104.

(49) Le, L.; Patterson, A.; White, M. Supervised Autoencoders: Improving Generalization Performance with Unsupervised Regularizers. *Adv. Neural Inf. Process. Syst.* **2018**, 107–117.

(50) https://scikit-learn.org/stable/.

(51) https://pytorch.org/.

(52) Fawcett, T. An Introduction to ROC Analysis. *Pattern Recognit. Lett.* **2006**, *27* (8), 861–874. https://doi.org/10.1016/j.patrec.2005.10.010.

(53) Śmieja, M.; Warszycki, D.; Tabor, J.; Bojarski, A. J. Asymmetric Clustering Index in a Case Study of 5-HT1A Receptor Ligands. *PLoS One* **2014**, *9* (7), e102069. https://doi.org/10.1371/journal.pone.0102069.

(54) Warszycki, D.; Śmieja, M.; Kafel, R. Practical Application of the Average Information Content Maximization (AIC-MAX) Algorithm: Selection of the Most Important Structural Features for Serotonin Receptor Ligands. *Mol. Divers.* **2017**, *21* (2), 407–412. https://doi.org/10.1007/s11030-017-9729-8.

Table of Contents graphic

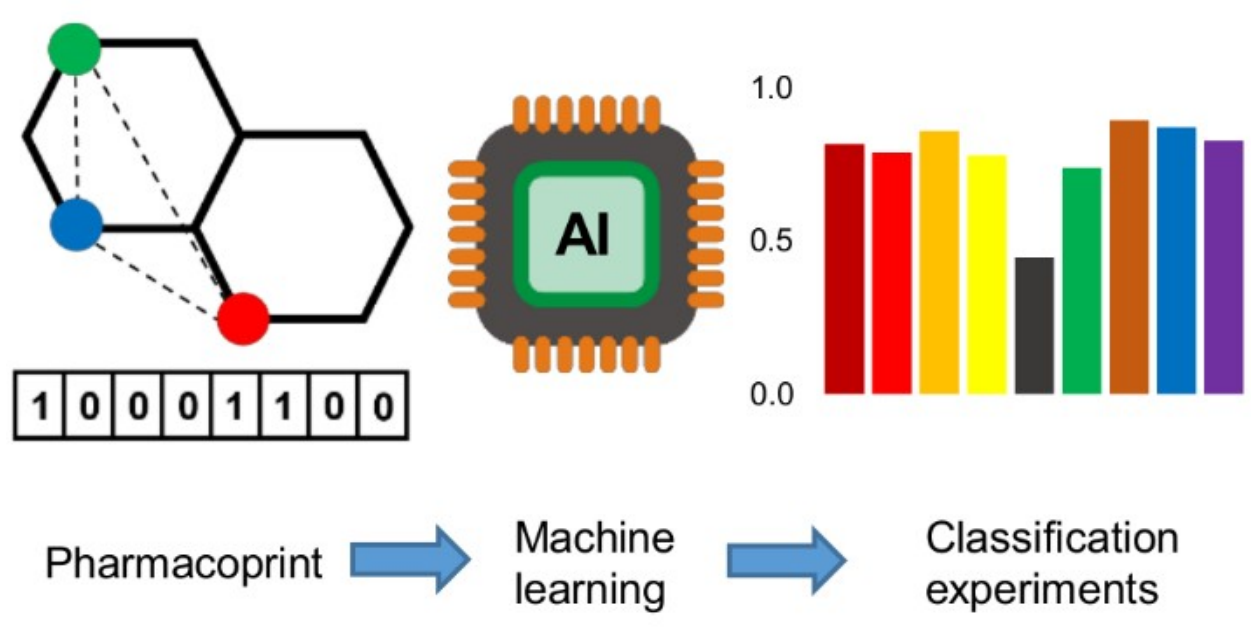